\documentclass[runningheads]{llncs}
\usepackage[T1]{fontenc}
\usepackage{xurl}
\usepackage{graphicx}
\usepackage[caption=false]{subfig}
\begin{document}
\title{\LARGE \bf
Assessing the Impact of Sanctions in the Crypto Ecosystem: Effective Measures or Ineffective Deterrents?
}
%
%
\author{}
\author{Francesco Zola\inst{1}\orcidID{0000-0002-1733-5515} \and Jon Ander Medina\inst{1}\orcidID{0009-0008-1107-0617} \and Raúl Orduna\inst{1}\orcidID{0000-0002-5932-0987}}
\titlerunning{Assessing the Impact of Sanctions in the Crypto Ecosystem}
\authorrunning{F. Zola et al.}
%
\institute{Vicomtech Foundation, Basque Research and Technology Alliance (BRTA);\\ Paseo Mikeletegi, 57, Donostia 20009, Spain \\
\email{\{fzola, jmedina, rorduna\}@vicomtech.org}}
\maketitle              
\begin{abstract}
Regulatory authorities aim to tackle illegal activities by targeting the economic incentives that drive such behaviour. This is typically achieved through the implementation of financial sanctions against the entities involved in the crimes. However, the rise of cryptocurrencies has presented new challenges, allowing entities to evade these sanctions and continue criminal operations. Consequently, enforcement measures have been expanded to include crypto assets information of sanctioned entities. Yet, due to the nature of the crypto ecosystem, blocking or freezing these digital assets is harder and, in some cases, such as with Bitcoin, unfeasible. Therefore, sanctions serve merely as deterrents. For this reason, in this study, we aim to assess the impact of these sanctions on entities' crypto activities, particularly those related to the Bitcoin ecosystem. Our objective is to shed light on the validity and effectiveness (or lack thereof) of such countermeasures. Specifically, we analyse the transactions and the amount of USD moved by punished entities that possess crypto addresses after being sanctioned by the authority agency. Results indicate that while sanctions have been effective for half of the examined entities, the others continue to move funds through sanctioned addresses. Furthermore, punished entities demonstrate a preference for utilising rapid exchange services to convert their funds, rather than employing dedicated money laundering services. To the best of our knowledge, this study offers valuable insights into how entities use crypto assets to circumvent sanctions.


\end{abstract}
\section{Introduction}

Understanding the multidimensional nature of crime is crucial for developing effective strategies to prevent and combat these illicit activities. Crimes often manifest in various forms and domains, like drug trafficking, human trafficking, and other types of organised crime. Nevertheless, in all these cases, the main goal of the criminal networks is still to make a profit \cite{neumann2017introduction}. For this reason, disrupting the economic incentives driving illicit behaviour has become the primary objective in tackling these crimes \cite{viscusi1986market}. This task requires cooperation and coordination among governments, law enforcement agencies, financial institutions, regulatory bodies, and other stakeholders at national and international levels.

This is the case of authority agencies such as the Office of Foreign Assets Control (OFAC)\footnote{https://ofac.treasury.gov/}, Office of Financial Sanctions Implementation (OFSI)\footnote{https://sanctionssearchapp.ofsi.hmtreasury.gov.uk/}, European External Action Service (EEAS)\footnote{https://www.eeas.europa.eu/} and United Nations Security Council (UNSC)\footnote{https://www.un.org/securitycouncil/content/un-sc-consolidated-list}, that aim to implement and enforce financial sanctions directly to the entity behind some violations and crimes. In fact, these agencies have the authority to freeze, block or restrict access to sanctioned entities' assets such as banking accounts, real estate, vessels, etc. 


However, with the advent of virtual currencies such as cryptocurrencies, stablecoins, and Non-Fungible Tokens (NFTs), sanctioned entities have discovered new opportunities for circumventing sanctions and continuing their illicit activities \cite{wardani2022money}. These digital assets promote decentralization and offer varying degrees of anonymity or pseudo-anonymity, creating a borderless ecosystem ideal for the proliferation of illicit activities \cite{connolly2019rise}\cite{toolCyber}. According to ``\textit{The 2024 Crypto Crime Report}'' \cite{chainalysis2024crypto}, although illicit crypto-transactions constitute less than 0.5\% of the total on-chain transaction volume, they accounted for nearly 40 billion USD in 2022 and over 24 billion USD in 2023.

The significant size of the crypto market, the opportunities crypto assets present, and their proven involvement in illicit activities \cite{bele2021cryptocurrencies}\cite{guidara2022cryptocurrency} have raised concerns about ensuring compliance with existing financial regulations. While some countries such as Tunisia, Nepal, Libya, Iraq, Bolivia, and Algeria have banned the use of cryptocurrency \cite{banned}, others have directed their effort to implement new frameworks and technology to increase their control degree over the crypto ecosystem. Thus, regulatory agencies have intensified their enforcement actions against sanctioned entities, including tracking information about their crypto assets whenever possible. However, due to the nature of the crypto ecosystem, they still face limitations in blocking these digital assets. As a result, sanctions solely serve as a deterrent, aiming to discourage other individuals and companies from engaging in transactions with punished entities. Yet, this limitation makes crypto assets the perfect facilitator for ongoing illicit operations and circumventing sanctions.

For this reason, in this work, we aim to examine the impact these sanctions generate in the crypto activities of punished entities and their related violations. Our objective is to evaluate the validity and effectiveness (or lack thereof) of such countermeasures. Specifically, we analyse the transactions and the amount of USD moved by punished entities that possess crypto addresses before and after being sanctioned by the authority agency. Furthermore, we investigate which type of known crypto entities (Exchange, Mixers, Gambling, etc.) are typically engaged with by the punished entities post-sanction. Thus, we investigate if they are attempting to connect with other entities involved in illicit operations or are employing strategies such as money laundering, on-ramps and off-ramps operations, funding raise campaigns, etc.

More specifically, this work analyses entities (individuals and companies) sanctioned by the OFAC agency that have information about Bitcoin (BTC) assets. These decisions were taken for two specific reasons: a) Bitcoin (together with stablecoins) is widely recognised among the most popular cryptocurrencies for illicit activities \cite{chainalysis2024crypto}, primarily due to its high market value and accessibility, even for users without technical background; b) other authority agencies (OFSI, EEAS, UNSC) do not have a comprehensive list of sanctioned crypto-related entities or do not release it publicly. 

The results suggest that sanctions have been effective for roughly half of the sanctioned entities, while the others continue to engage in transactions through sanctioned Bitcoin addresses. Additionally, sanctioned entities prefer to directly convert their cryptocurrencies using dedicated services (\textit{Exchanges}), rather than apply money laundering strategies using services like \textit{Mixers} or \textit{Gambling}.


To the best of our knowledge, this study offers a first step towards determining the effectiveness of sanctions within the crypto ecosystem and how sanctioned entities used them to circumvent sanctions.

\section{Background} \label{sec:related}
This section presents an overview of the crypto ecosystem, showing regulations, directives, and literature approaches. Specifically, in Section \ref{sec:policy} a review of European Union regulations related to the crypto ecosystem is presented, while Section \ref{sec:ofac} is focused on presenting the United States OFAC agency and its crypto sanctions. Finally, Section \ref{sub:sota} describes related research in the field.

\subsection{EU Directive review} \label{sec:policy}

As mentioned, crypto assets can be harder (or in some cases unfeasible) to freeze or block directly. However, governments and regulatory authorities worldwide have been increasingly implementing measures to monitor and regulate crypto markets to tackle concerns related to illicit activities. Thus, these regulations aim to facilitate the imposition of sanctions when necessary. 

The European Union (EU) has established and reviewed several directives aimed at tackling issues such as money laundering, fraud, terrorism financing, and other emerging challenges related to non-cash payments. For instance, in 2015, the 5th Anti-Money Laundering Directive (5AMLD) \cite{2015/849} was introduced to address new trends in terrorist financing, building upon the provisions of the 4AMLD. Notably, under the 5AMLD, the role of cryptocurrency Exchanges has changed, they are now considered equivalent to financial institutions. Therefore, they are required, among other measures, to adhere to Know Your Customer (KYC) requirements, implement Anti-Money Laundering (AML) mechanisms, and register with national regulatory authorities. This directive has been amended to include provisions regarding information accompanying transfers of funds and certain crypto-assets, through the Regulation (EU) 2023/1113 \cite{2023/1113}. The amendment fosters international cooperation within the Financial Action Task Force (FATF) and the global implementation of its recommendations \cite{FATF}. Furthermore, it sets the obligation for virtual asset service providers (VASPs) and crypto asset service providers (CASPs) to collect information about the person who uses their services. Together with these directives, another pivotal regulation is the Markets in Crypto-Assets (MiCA) \cite{mica2023}, which clearly distinguishes between different types of crypto-assets (asset-referenced tokens, electronic money or e-money, and other crypto assets). It then imposes constraints on CASPs to ensure market integrity and financial stability. One of its key provisions involves significant disclosure and transparency rules aimed at better informing consumers about associated risks, as well as mandating the implementation of security measures and anti-money laundering compliance.

Fraud and counterfeiting of non-cash means of payment is regulated through the EU directive 2019/713 \cite{2019/713}. The framework establishes measures aimed at preventing and detecting fraud and counterfeiting of non-cash payment instruments, such as security requirements for payment service providers, customer authentication procedures, and usage of secure technologies (encryption and tokenization). This directive aimed to cover also new types of non-cash payment instruments such as e-money and virtual currencies, since they have a significant cross-border dimension. Another interesting EU framework, although it doesn't specifically mention digital currency or cryptocurrencies due to its general aim, is the Directive (EU) 2017/1371 \cite{2017/1371} that defines legal framework and measures to combat any fraud against the financial interests of EU. 

Despite the revision, update, and implementation of these policies, freezing or blocking crypto assets remains harder to accomplish by technical design. Its boundary-less structure, the availability of services in jurisdictions where these policies don't apply, or in countries unwilling to cooperate in criminal investigations, the anonymity of users, and the ease of conducting transactions - these properties collectively enable users to circumvent sanctions and persist in their illicit activities.

\subsection{US Office of Foreign Assets Control}\label{sec:ofac}

The Office of Foreign Assets Control (OFAC) is part of the Department of the United States (US) Treasury, and it is in charge of implementing economic and trade sanctions in accordance with US foreign policy and national security objectives. These sanctions are directed towards specific foreign countries and regimes, terrorists, international narcotics traffickers, individuals involved in the proliferation of weapons of mass destruction, and other actors posing threats to the national security, foreign policy, or economy of the United States. These actors and their blocked assets are included in a \textit{Specially Designated Nationals and Blocked Persons} (SDN) list\footnote{https://sanctionssearch.ofac.treas.gov/}. Consequently, US individuals and companies are generally prohibited from dealing with them. The actors reported in the SDN list, which we refer to as \textit{entities} in this paper, are sanctioned due to the violation of one (or more) Executive Orders and/or Code of Federal Regulations (CFR)\footnote{https://ofac.treasury.gov/specially-designated-nationals-list-sdn-list/program-tag-definitions-for-ofac-sanctions-lists}. Since 2018, OFAC has included cryptocurrency-related information in the SDN list as blocked assets for sanctioned entities whenever such information is available. In some cases, one entity can have multiple sanctioned addresses. To date, the violations that have led to sanctions against entities with crypto addresses involved, are detailed in Table \ref{tab:sanctionlists}.
In this paper

\begin{table}
\resizebox{\linewidth}{!}{
\centering
\begin{tabular}{llll}
\hline
\textbf{\#} &\textbf{Code} &\textbf{Description} & \textbf{Executive Order N.} \\\hline
1 &CYBER2 &Malicious Cyber Activities & 13694, 13757   \\
2 &DPKR3 &Blocking Property of the Government of North Korea &13722 \\
3 &DPKR4 &Additional Sanctions With Respect to North Korea &13810 \\
4 &ELECTION &Foreign Interference in the US Election &13848 \\
5 &IFSR &Iranian Financial Sanctions Regulations &31 CFR part 561 \\
6 &ILLICTI-DRUGS &Illicit Drug Trade &14059 \\
7 &IRGC &Iranian Financial Sanctions &31 CFR Part 561 \\
8 &NPWMD &Weapons of Mass Destruction Proliferators Sanctions &31 CFR part 544 \\
9 &RUSSIA &Blocking Harmful Activities of the Russian Federation  &14024 \\
10 &SDGT &Narcotics Trafficking Sanctions &  31 CFR part 594 \\
11 &SDNTK &Foreign Narcotics Kingpin Sanctions &31 CFR part 598 \\
\hline
\end{tabular}}
\caption{Violations reported in the SDN list that have generated sanctions against cryptocurrency related entities.}
\label{tab:sanctionlists}
\end{table}

\subsection{Cryptocurrency and Cybercrime} \label{sub:sota}

Cryptocurrencies have created a convenient ecosystem for the permanence and movement of illicit cybercrime-related activities, with currencies such as Bitcoin, Ethereum, and Monero becoming their operational space. The evolution of illicit activities within the cryptocurrency ecosystem has been studied extensively \cite{cole2024virtual} \cite{blanchini2022supporting}. For instance, Hornuf et al. \cite{hornuf2023cybercrime} analysed cybercrime related to Ethereum transactions. In particular, they identified over 1.78 million transactions related to 19 categories of cybercrime, with losses amounting to \$1.65 billion up to the year 2021, posing the focus on estimating how these cybercrimes impact victims’ risk-taking, risk-adjusted returns, and investor behaviour. In the same line, in \cite{sanusi2022cryptocurrency}, authors employ the Generalized Autoregressive Score (GAS) model to examine the impact of cybercrime on cryptocurrency returns in South Africa. On the other hand, in \cite{10039632}, authors attempted to correlate the expansion of ransomware activities with transactions performed in these cryptocurrencies between 2015 and 2020. However, although it is clear that these cryptocurrencies facilitate the growth of ransomware revenue \cite{chainalysis2024crypto}, authors did not find an evident correlation.  In \cite{alotibi2022money}, traditional machine learning algorithms are used to detect illegal activities using a Bitcoin dataset, while in \cite{japinye2024integrating} graph-based networks are used for a similar task. Similar applications are also explored in \cite{10.1145/3589334.3645692} for defining a method to identify and trace illicit activities in the Ethereum blockchain.

Among the most relevant cybercrimes, cryptocurrencies have become the perfect solution for laundering illegal funds \cite{soudijn2024encounters}. By fostering decentralization, user anonymity, and the ease of making cross-border transactions, they have led to the proliferation of dedicated services. Achraf Guidara \cite{guidara2022cryptocurrency} examines the relationship between cryptocurrencies and money laundering, emphasising the urgent need to develop a robust and internationally coordinated regulatory framework to mitigate these risks. In \cite{nazzari2023payday}, 182 Bitcoin addresses belonging to 56 members of the Conti ransomware group are analysed with the aim of identifying if money laundering mechanisms are applied. They conclude that cryptocurrency exchanges and dark web services are involved in 71\% and 30\% of transactions, respectively, while only 8\% utilized mixers. These findings challenge the prevailing notion that cybercriminals employ sophisticated methods, highlighting instead the simplicity of their tactics \cite{nazzari2023lost}.

At the same time, as introduced in the previous sections, the lack of market control has turned these cryptocurrencies into a means of evading sanctions, allowing entities to continue their illicit activities. For this reason, in this work, we aim to analyse how punished entities use crypto assets to circumvent sanctions and whether they also employ laundering mechanisms to increase their anonymity.
 
\section{Experimental Framework} \label{sec:study}
In this section, the dataset and the approach followed in this study are presented. More specifically, Section \ref{sec:sanctions} reviews the sanctioned list used, while Section \ref{subsec:dataset} introduces the Bitcoin dataset. Finally, the guidelines and key concepts used during the experiments are reported in Section \ref{subsec:graph}.

\subsection{Sanctions List}\label{sec:sanctions}
\begin{figure}[!htbp]
  \centering
  \subfloat[Distribution of sanctioned cryptocurrencies in the SDN list.\label{fig:distribution_crypto}]{%
    \includegraphics[width=0.50\linewidth]{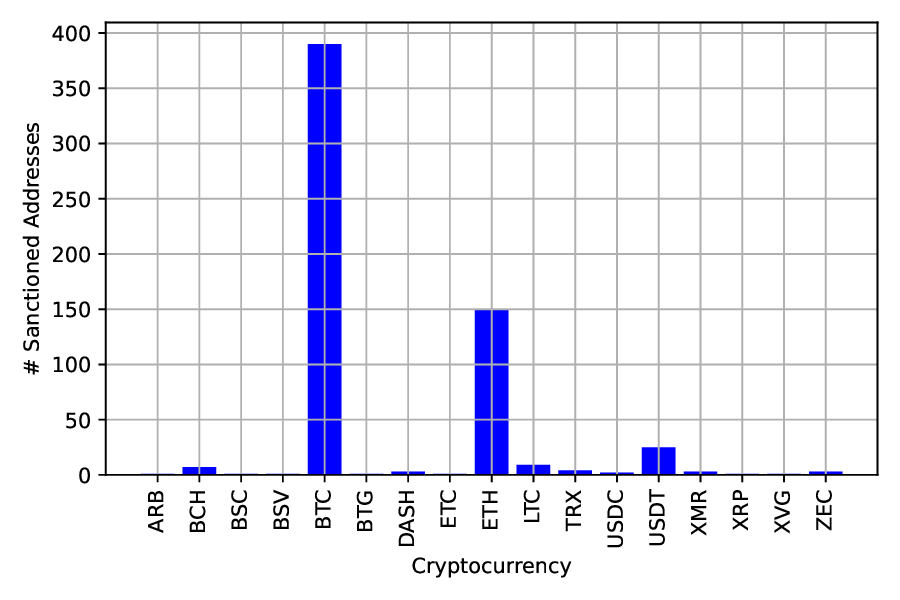}
    }\hfill
 \subfloat[Addresses and entities distribution in the SDN list (top-5 populated cryptocurrencies).\label{fig:company}]{%
    \includegraphics[width=0.48\linewidth]{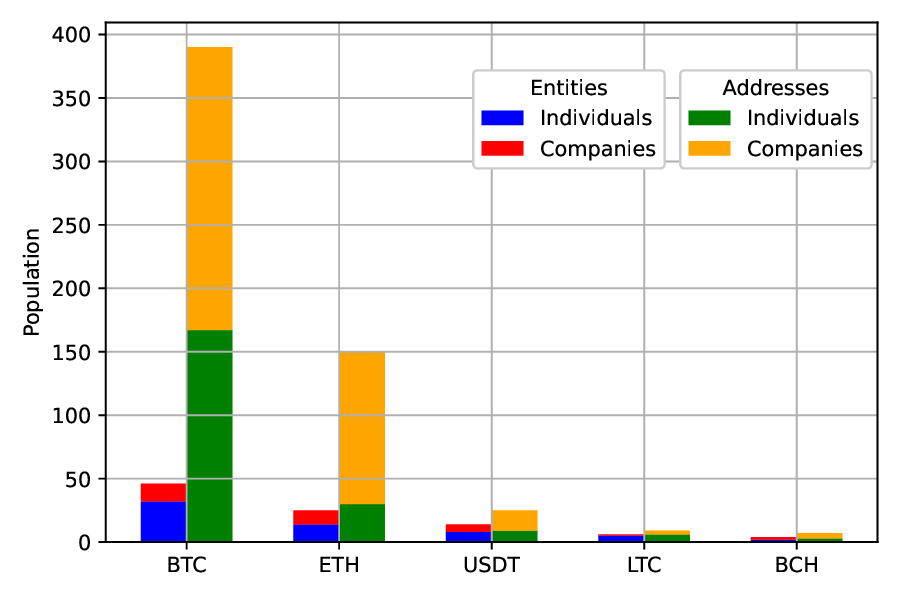}}
  \caption{Overall statistics of sanctioned entities with cryptocurrency information extracted from the SDN list (February 2024).}
  \label{fig:dataset_hist1}
\end{figure}

As of February 2024, the SDN list contains information about 600 crypto addresses related to 17 different cryptocurrencies, as shown in Figure \ref{fig:distribution_crypto}. The figure shows that the majority of reported addresses ($\sim 65$\%) belong to the Bitcoin (BTC) network, while another $\sim 25$\% are from Ethereum (ETH). The remaining addresses, representing just 10\%, are divided across 15 cryptocurrencies. Furthermore, analysing the composition of the sanctioned entities (Figure \ref{fig:company}) that belong to the top-5 sanctioned cryptocurrencies, it becomes evident that while there are more punished individuals than companies for both BTC and ETH, companies possess a greater number of punished addresses. These results lead us to focus the analysis only on the BTC-related entities, as anticipated in Section \ref{sec:ofac}. In particular, this constraint leaves us to study 43 out of the 56 available entities in the SDN list.

\begin{figure}[!htbp]
  \centering
 \subfloat[Distribution of BTC-related entities in the SDN list per country/region.\label{fig:country}]{%
    \includegraphics[width=0.49\linewidth]{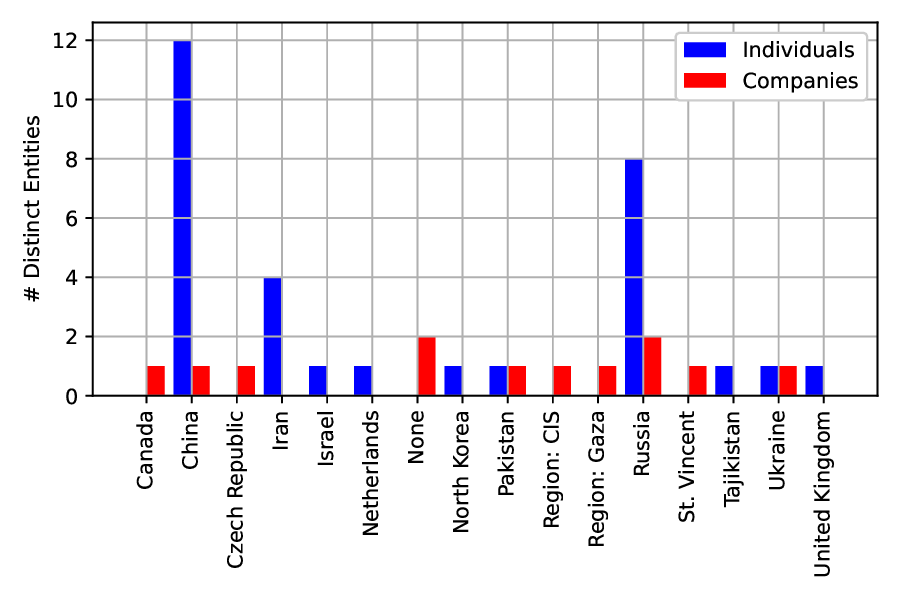}}\hfill
 \subfloat[Distribution of BTC-related entities in the SDN list per violation.\label{fig:code}]{%
    \includegraphics[width=0.49\linewidth]{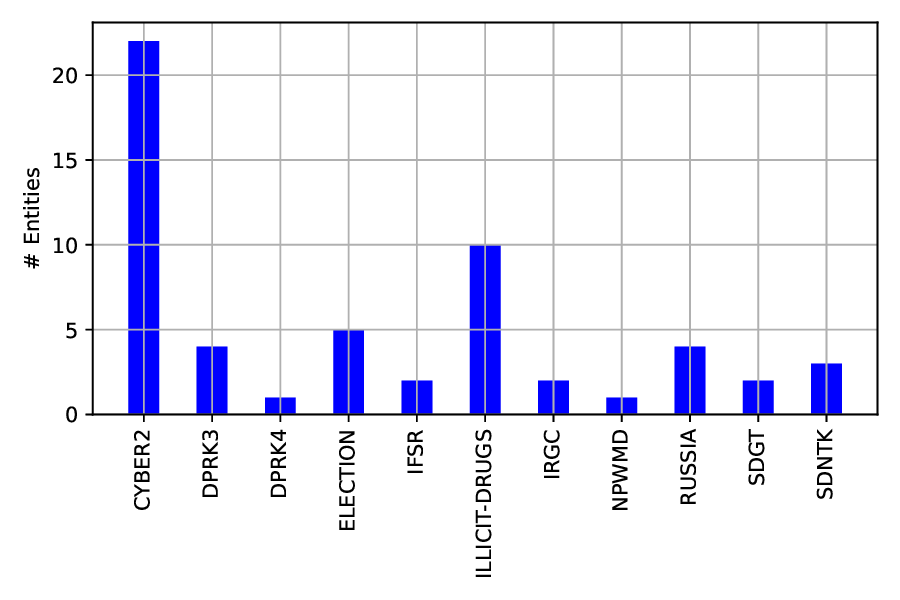}}\hfill
 \subfloat[BTC-related entities per violation over time.\label{fig:years}]{%
   \includegraphics[width=0.50\linewidth]{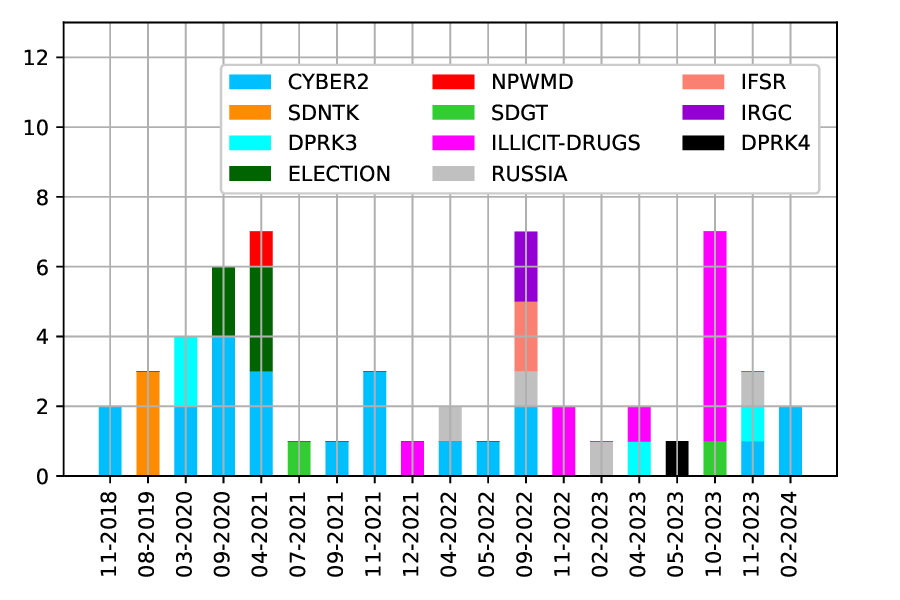}}
  \caption{Overview about violations of sanctioned entities with BTC information extracted from the SDN list (February 2024).}
  \label{fig:dataset_hist2}
\end{figure}

Figure \ref{fig:country} details that most of the BTC-related sanctioned entities (both companies and individuals) are located in China and Russia. In fact, of the 43 punished entities, 13 are from China and 10 from Russia ($\sim54$\%). In both cases, the number of sanctioned individuals overwhelms the number of companies, while there are only sanctioned companies in Canada, the Czech Republic, St. Vincent, and in the regions of Gaza and the Commonwealth of Independent States (CIS). Figure \ref{fig:code} shows the distribution of these sanctioned entities with respect to their violations. It is to be noted that one entity can face sanctions for multiple violations. Consequently, the most prevalent category, with 23 entities, pertains to malicious cyber-enabled activities since they include a broad spectrum of crimes. Furthermore, among the most populated categories, 10 entities are sanctioned for illicit drug trading, while 5 entities are linked to interference in the US election. Finally, Figure \ref{fig:years} analyses the temporality of these sanctions and the number of entities involved. The figure shows that US authorities led a significant operation on illicit drug trading in October 2023, resulting in the sanctioning of 6 entities. Another interesting finding is that sanctioning operations against specific violations tend to occur only on specific dates. Specifically, violations related to \textit{CYBER2}, \textit{ILLICIT-DRUGS}, and \textit{RUSSIA} are consistently detected over time, while the others are detected only on specific dates.

\subsection{Bitcoin Dataset}\label{subsec:dataset}
In this paper, the entire Bitcoin blockchain data until the block 830,000 are downloaded, i.e., all the transactions until February 11th, 2024 (more than 900M transactions). On the other hand, to have more information about real-world entities, labelled (tagged) addresses are gathered from multiple reliable sources, such as WalletExplorer\footnote{https://www.walletexplorer.com/} and the tagpacks provided by Graphsense\footnote{https://graphsense.info/}. Indeed, these sources represent valid solutions used in many previous researches \cite{paquet2019ransomware}\cite{tovanich2023fingerprinting}\cite{zola2019cascading}, and allowed us to gather more than 38M addresses of almost 400 entities labelled as \textit{Exchanges, Gambling, Marketplaces, Mining Pools, Mixers, Services, Trading platforms, eWallet, Ransomware, Sextortion,} and \textit{Extremist}.

\subsection{Proposed Analysis}\label{subsec:graph}
\begin{figure}[!htbp]
  \centering
    \includegraphics[width=0.6\linewidth]{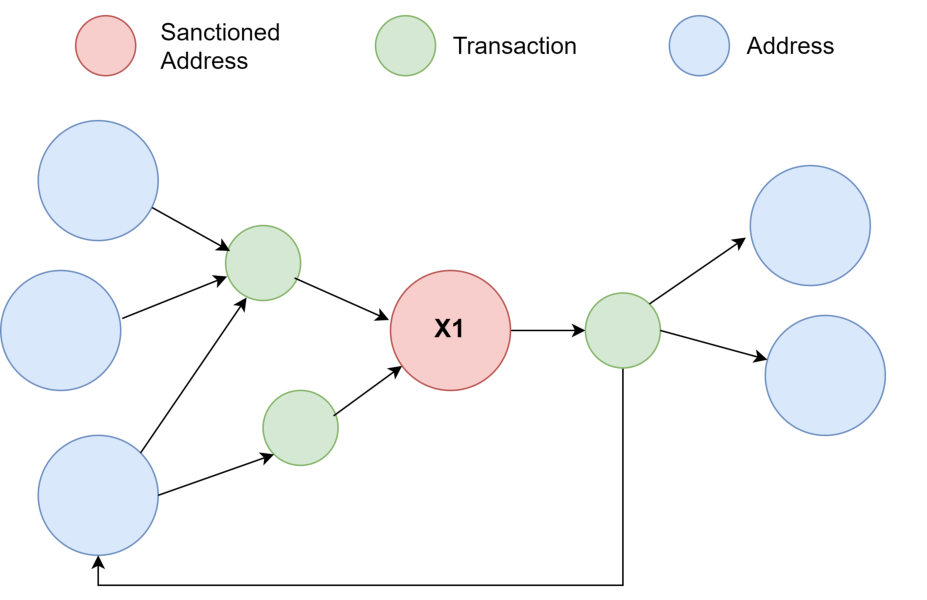}
  \caption{An example of a \textit{1}-step address-transaction graph}
  \label{fig:addresstransaction}
\end{figure}

As mentioned in Section \ref{sec:sanctions}, the BTC addresses included in the SDN list represent the starting point of our investigation. More specifically, from each of them, we analyse the address-transaction graph \cite{fleder2015bitcoin}\cite{zola2019cascading}. This graph is directly built using the information available in the BTC blockchain, where nodes are BTC addresses and transactions. Then directed edges (arrows) from addresses to transactions represent incoming relations, while edges from transactions to addresses are outgoing relations, as shown in Figure \ref{fig:addresstransaction}. Furthermore, the edge may incorporate BTC information like amount, fee, timestamps, etc. With these principles, it is possible to define the $n$-step address-transaction graph of a sanctioned address $X1$, as a graph in which all the paths from $X1$ involve maximum $n$ transactions. Thus, the paths from $X1$ have a maximum length of $2n$ (Figure \ref{fig:addresstransaction}).

In this work, we present two different analyses: the first one is based on assessing the effectiveness of the sanctions by analysing the activities of the entities (\textit{flow analysis}), and the second has the aim to detect the relations that entities have after being sanctioned (\textit{behavioural analysis}).

For the \textit{flow analysis}, a temporal aspect is introduced in the address-transaction graph. Specifically, for each address of each entity, multiple 1-step address-transaction graphs are created, considering 4 different temporal ranges: a) all the transactions prior to the imposed sanction (\textit{pre-sanction}); b) transactions achieved immediately after the sanctions within the subsequent 7 days (\textit{7 post-sanction}); c) transactions achieved immediately after the sanctions within 30 days post-sanctions (\textit{30 post-sanction}); d) and finally all the activities post-sanctions up to February 11th, 2024 (\textit{up-to-date}). 
These ranges allow us to evaluate the behaviour of the sanctioned entity and detect how they react to this situation in short, medium, and long terms. 

Once these graphs are built, from each one, several metrics such as the number of input and output transactions, the overall balance of the entity after each temporal range, and the amount in USD of money sent and received by the entity (the BTC/USD value is fixed on the day the transaction is performed), are extracted and used for evaluating the trends and the effectiveness of the sanctions. It is to be noted that one entity may possess multiple sanctioned addresses. Therefore, metrics computed from each of its addresses are aggregated to provide a comprehensive overview of the entity's behaviour.

On the other hand, the \textit{behavioural analysis} is based on analysing a single address-transaction graph for each entity, that is created using data from immediately after the sanctions until the end of the dataset (\textit{up-to-date}). Furthermore, this graph is enriched with real-world entity information, i.e., with labels gathered from the external sources mentioned in Section \ref{subsec:dataset}. This approach enables us to identify whether the sanctioned entity engages in transactions with other known entities, which could include potential actors related to illicit operations (e.g. other sanctioned entities, ransomware, etc.) or it tries to apply strategies for conducting activities such as money laundering (e.g. involving mainly mixers, gambling or other services), on-ramps and off-ramps operations (e.g. involving exchanges), funding raise campaign (e.g. involving mining pool or marketplace). This \textit{behavioural analysis} is performed considering both 1-step and 2-step address-transaction graphs. This approach gives us a deeper view of the entity strategy, since the 1-step analysis only provides information about its direct relations, while the 2-step also includes undirected transactions (reached in two steps).

\section{Current Study} \label{sec:study}
In this section, the results obtained in the two proposed analyses are presented. In particular, Section \ref{subsec:flow} describes the results obtained analysing transactions and money flow of the entities before and after their sanctions, while Section \ref{subsec:behaviour} details the relations that the sanctioned entities have had with other known type of entities. Finally, discussions and limitations are reported in Section \ref{subsec:discussion}.

\subsection{Flow Analysis Results}\label{subsec:flow}
\begin{figure}[!htbp]
  \centering
    \includegraphics[width=0.6\linewidth]{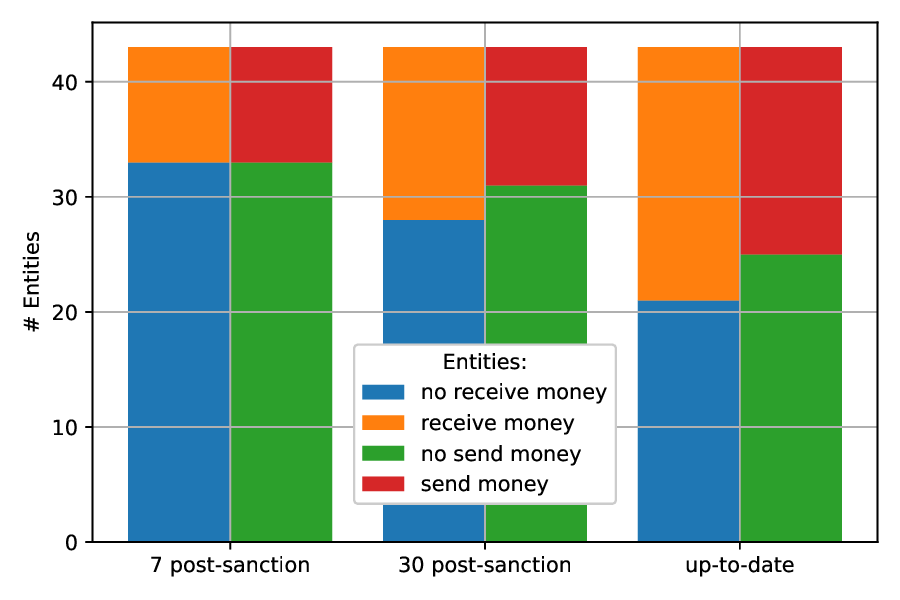}
  \caption{Number of sanctioned entities that perform transactions (received and sent) in the different post-sanction intervals.}
  \label{fig:transactiontrend}
\end{figure}

\begin{figure}[!htbp]
  \centering
    \includegraphics[width=0.6\linewidth]{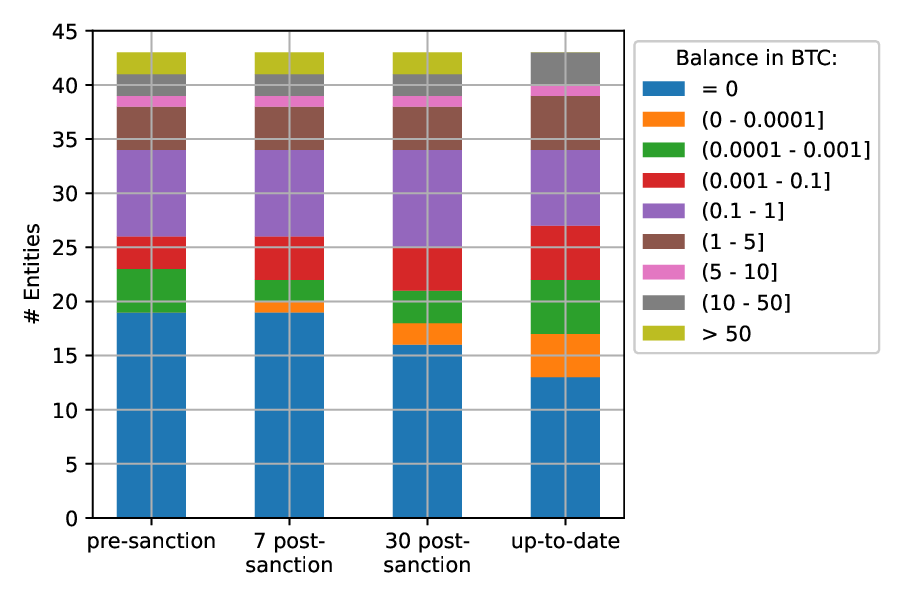}
  \caption{Entity balance (in BTC) considering the sanctioned addresses at the four stages: \textit{pre-sanction, 7 post-sanction, 30 post-sanction,} and \textit{up-to-date}.}
  \label{fig:balancetrend}
\end{figure}

Figure \ref{fig:transactiontrend} shows the number of entities that received and sent money through crypto transactions post-sanctions. Specifically, the figure illustrates that only half of all the sanctioned entities were effectively discouraged from engaging in transactions. In particular, only 21 entities stopped to receive money, and 25 to send funds. On the other hand, the figure reveals that despite the sanctions, some entities (7) continued to move funds within 7 days of the OFAC sanction. For this reason, Figure \ref{fig:balancetrend} enables us to comprehend how these funds are being moved, analysing the balance in terms of BTC held by each entity in its sanctioned addresses before and after the sanctions.


Although Figure \ref{fig:transactiontrend} indicates that some entities were not deterred from conducting transactions, Figure \ref{fig:balancetrend} emphasises a general trend of maintaining at least a minimal balance in the sanctioned addresses, excluding the two entities with the highest balance ($\geq 50$ BTC) who adopted an off-ramp strategy. Yet, the number of entities with a balance of 0 decreased, while the number of entities with a balance in a range $>$ 0 and $\leq$ 0.1 BTC increased.

Table \ref{fig:volumeUSD} reports the number of transactions that involve sanctioned entities, categorised by the violation they are convicted of. Additionally, the table includes the USD volume moved by the entities for each violation, both before and after the sanctions. In particular, the volume indicates both incoming and outgoing transactions. Results in Table \ref{fig:volumeUSD} show that \textit{CYBER2} represents the violation with the highest number of transactions pre-sanction (more than 150K) and the highest USD volume of about 8,300 million. The result is expected since this violation also has the highest number of sanctioned entities (Figure \ref{fig:code}). However, what is interesting, is that after the sanctions, entities related to this violation still perform 305 transactions for a market of about 3 million USD. On the other hand, entities related to \textit{DPRK3} and \textit{ILLICIT-DRUGS} violations perform a high number of transactions and show a USD volume of 240 million and 120 million pre-sanction, respectively. Although these numbers decrease in the post-sanction phase, they remain consistent, with 209 thousand USD for \textit{DPRK3} and 8 million USD for \textit{ILLICIT-DRUGS}. Also, the market related to \textit{RUSSIA} violation is pretty high pre-sanctions, with almost 40 million USD moved in just 553 transactions. However, it seems strongly affected by the sanctions, resulting in just 4 transactions with an overall amount of 162 USD. Finally, entities related to violations such as \textit{IFSR, IRGC} and \textit{SDGT} are indeed deterred from performing transactions; in fact, they achieve only 1 transaction post-sanctions, moving just a few dollars (or less). Notably, the entities involved in \textit{DPRK4} activities are shut down. More precisely, they have not achieved transactions from May 2023 (Figure \ref{fig:years}) until the date.

\begin{table}[]
\centering
\begin{tabular}{ll|rr|rr}
\hline
 &  & \multicolumn{2}{c|}{\textbf{\# Transactions}} & \multicolumn{2}{c}{\textbf{USD Volume}} \\ \hline
\multicolumn{1}{c}{\textbf{\#}} & \multicolumn{1}{c|}{\textbf{Violation}} & \textbf{Pre-Sanction} & \textbf{Up-to-date} & \multicolumn{1}{c}{\textbf{Pre-Sanction}} & \multicolumn{1}{c}{\textbf{Up-to-date}} \\ \hline

1 & \textit{CYBER2} &153 K  &305    &8,300 M  &3 M    \\
2 & \textit{DPRK3} &2 K  &63    &240 M  &209 K    \\
3 & \textit{DPRK4} &62  &0  &10 M   &0    \\
4 & \textit{ELECTION} &46 K  &8 &6 M  &1 K    \\
5 & \textit{IFSR} &182  &1  &461 K  &$\leq1$  \\
6 & \textit{ILLICIT-DRUGS} &9 K  &747  &120 M  &8 M \\
7 & \textit{IRGC} &182  &1    &461 K &$\leq1$  \\
8 & \textit{NPWMD} &97  &8  &26 K  &1 K   \\
9 & \textit{RUSSIA} &553  &4   &39 M  &162   \\
10 & \textit{SDGT} &18 K  &1   &42 M  &103    \\
11 & \textit{SDNTK} &354  &42   &104 K  &12 K  \\\hline

\end{tabular}
\caption{Number of transactions achieved and estimation of USD moved before and after the sanctions for each violation.}
 \label{fig:volumeUSD}
\end{table}

\subsection{Behavioural Analysis Results}\label{subsec:behaviour}

Table \ref{tab:stepanalysis} reports the results gathered during the behavioural analysis, involving 1-step and 2-step address-transaction graphs. In particular, by creating a 1-step graph from each of the 43 entities (387 BTC-sanctioned addresses) it is possible to reach about 4K addresses, of which only 340 addresses (8.48\%) are related to known entities and have an external label. On the other hand, increasing the analysis considering also undirected connections (2-step address-transaction graphs), it is possible to reach more than 10M addresses, of which just 175,444 (1.67\%) are labelled. These outcomes confirm that, although the 1-step analysis has more labelled information, it identifies only 15 known entities of 4 different behaviours, while the 2-step analysis enriches the investigation by uncovering 67 entities with 9 different behaviours. 

The results reported in Table \ref{tab:stepanalysis} highlight that, in both the 1-step and 2-step analyses, the majority of labelled addresses belonged to \textit{Exchanges} with 97\% and 86\%, respectively. At the same time, in the 2-step analysis, 13.5\% of the labelled addresses are linked to 11 crypto services (trading, eWallet, banking, etc.). This enhanced scenario also highlights connections between sanctioned entities and addresses associated with \textit{Sextortion, Ransomware} and \textit{Extremism} crimes. 

\begin{table}[]
\centering\begin{tabular}{ll|cc|cc}
\hline
 &  & \multicolumn{2}{c|}{\textbf{1-step analysis}} & \multicolumn{2}{c}{\textbf{2-step analysis}} \\\hline
\multicolumn{1}{c}{\textbf{\#}} & \multicolumn{1}{c|}{\textbf{\begin{tabular}[c]{@{}c@{}}Behaviour\end{tabular}}} & \textbf{\begin{tabular}[c]{@{}c@{}}\# Distinct \\ Entities\end{tabular}}  & \textbf{\begin{tabular}[c]{@{}c@{}}\# Distinct \\ Address\end{tabular}} & \multicolumn{1}{c}{\textbf{\begin{tabular}[c]{@{}c@{}}\# Distinct \\ Entities\end{tabular}}} & \multicolumn{1}{c}{\textbf{\begin{tabular}[c]{@{}c@{}}\# Distinct \\ Address\end{tabular}}} \\\hline
1 & \textit{Service} & 3 & 5 &13  &23,727  \\
2 & \textit{Mixer} & 1 & 1 &2  &14  \\
3 & \textit{Exchange} & 9 & 331 &37  &151,385  \\
4 & \textit{OFAC Sanctioned} & 2 & 3 &6  &33  \\
5 & \textit{Gambling} & - & - &1  &249  \\ 
6 & \textit{Extremism} & - & - &1  &13  \\ 
7 & \textit{Ransomware} & - & - &2  &9  \\ 
8 & \textit{Mining Pool} & - & - &4  &13  \\ 
9 & \textit{Sextortion} & - & - &1  &1  \\ \hline
 & \textbf{Labelled} & \textbf{15} & \textbf{340} & \textbf{67} & \textbf{175,444} \\
 & \textbf{No Labelled} &\textbf{-} & \textbf{3,668} & \textbf{-} & \textbf{10,356,787} \\ \hline

\end{tabular}
\caption{Behavioural analysis of output entities related to sanctioned actors considering 1-step and 2-step address-transaction graphs.}
\label{tab:stepanalysis}
\end{table}

\subsection{Discussion and Limitations} \label{subsec:discussion}

\textbf{Discussion.} New regulations have started to treat Exchanges as financial services, requiring any Crypto Asset Service Provider or Virtual Asset Service Provider to implement anti-money laundering control measures such as Know Your Customer (KYC) policies. Nevertheless, criminals have started to explore new methods to obtain money fraudulently and to launder and use it, e.g., including dedicated services and other digital assets that current solutions/directives do not adequately supervise. Indeed, these new money laundering methods exploit gaps in existing legislations, which in turn require frequent updates and make them challenging to enforce and follow. In this context, it is crucial to consider and incorporate also technical connectivity to the crypto ecosystem, i.e., ensuring the technology used by both users and service providers is connected and operates under unified legal standards. This alignment will help enforce the law and catch those who commit crimes, making it more difficult for criminals to exploit any legal loopholes.

Aligned with the findings presented in \cite{soudijn2024encounters}, this study, using OFAC information, shows that Bitcoin is among the most used cryptocurrencies for various crimes, not limited to the cyber ecosystem. The results indicate that, despite being sanctioned, entities still perform operations with their blocked or frozen cryptocurrencies without employing complex tactics or dedicated money laundering services in their transactions. In fact, entities prefer to achieve off-ramp activities through known and reliable exchanges. These results are aligned with the outcomes reported in \cite{nazzari2023payday} and \cite{nazzari2023lost} regarding ransomware funds. However, the results also show that, in some cases, sanctioned entities maintain relationships with other sanctioned entities or entities involved in other crimes, such as sextortion, ransomware, and extremism.

This paper shows that a 1-step analysis is not sufficient for understanding and tracing the operations of sanctioned entities. Indeed, the analysis reveals that only a few sanctioned entities can be traced to known services, while a 2-step analysis provides enhanced context to their operations. However, although expanding the analysis to include more steps seems beneficial, it should be noted that this approach also introduces more unlabelled addresses, increasing the uncertainty in the "follow-the-money" investigation. Therefore, the number of steps to be included must be determined on a case-by-case basis.



\textbf{Limitation.} When interpreting the results, it is important to take into account some assumptions/constraints considered during this investigation. Firstly, the results of the \textit{behavioural analysis} strongly depend on the quantity and quality of labelled data available in the literature. The outcomes of this work are strictly related to the information provided by the US OFAC authority and need to be verified when information from new authorities becomes available. At the same time, regarding the quantity of the data, as introduced in Section \ref{subsec:dataset}, this work has gathered 38M addresses of almost 400 entities used in many state-of-the-art works \cite{paquet2019ransomware}\cite{tovanich2023fingerprinting}\cite{zola2019cascading}. Yet, these 38M represent only 2.9\% of the addresses used in the BTC blockchain, which counts about 1,300M addresses as of February 2024. Additionally, the OFAC SDN list being considered includes 40 sanctioned entities with Bitcoin addresses involved. While one might argue that this number is insufficient for comprehensive trend analysis, it should be noted that this is the maximum number available in the real ecosystem. On the other hand, regarding the quality of the information, we rely on the fact that the labelled datasets are used in many research investigations, as mentioned in section \ref{subsec:dataset}, and in some cases, they are extracted by tools that are used by LEAs in real investigations \cite{graphsense}. Moreover, although these datasets are generated and gathered from different sources, they do not present inconsistencies among them in the provided data.

Furthermore, it is to be noted that some entities were sanctioned in late 2023 or early 2024, meaning that the gathered crypto transactions (until February 2024) might not fully capture their activities. However, we decided to focus the analysis on the type of violation rather than concrete entities (Table \ref{fig:volumeUSD}). In fact, looking at the distribution analysis provided in Figure \ref{fig:years}, it is possible to see that the majority of the entities for each violation are prior to May 2023 - excluding the \textit{ILLICIT-DRUGS} crime - allowing us to analyse 9 months of transactions.

Finally, in the context of the 2-step analysis, this study has assumed that there has been no change in ownership of the funds. However, relying just on blockchain information, the validity of this assumption cannot be ensured. Nonetheless, we acknowledge this risk because limiting the analysis solely to 1-step graphs would be excessively restrictive, generating a skewed and poor view of the impact of the sanctions. Furthermore, a change in the ownership of the funds should not generate high-impact deviations in the proposed analysis, since we are just analysing the reached entities using a basic "follow-the-money" approach.

\section{Conclusion}\label{sec:conclusion}

The present study represents a first step and provides an interesting yet partial understanding of the impact of sanctions on the crypto ecosystem, with a focus on the Bitcoin cryptocurrency. The first point to emphasise is that sanctions are inherently tied to the prevailing regulations at any given time. As context, this study solely reports on European policies aimed at regulating the crypto market and preventing financial fraud. However, the analysis is conducted using data provided by the US OFAC authority, as it is the only entity that releases a comprehensive list of economically and trade-sanctioned entities.

The \textit{flow analysis} shows that in general, sanctions have been effective on at least half of the sanctioned entities, while the other half has continued to move (receive and send) money through the sanctioned BTC addresses, although they do not show huge changes in their current balance. In particular, sanctions seem to be not very effective against entities related to specific violations, like \textit{CYBER2} and \textit{ILLICIT-DRUGS} which are the ones that still make transactions and move high quantities of USD in proportion with their activities pre-sanctions. 
On the other hand, the \textit{behavioural analysis} highlights that sanctioned entities tend to prefer reaching out directly to \textit{Exchanges} to convert their cryptos, rather than use dedicated services for money laundering (mixers or gambling).

With the aim of deepening that understanding, further analysis should include heuristics assumptions in the loop \cite{zhang2020heuristic} as well as automatic labelling strategies to generate clustered entities. Yet, in this way, it would be possible for each punished entity, not only to consider the actual sanctioned addresses reported in the list, but also other addresses that are likely to belong to the same wallet. At the same time, it will be interesting to scale up our approach by incorporating information from other cryptocurrencies (Ethereum) and stablecoins. In fact, as reported in \cite{chainalysis2024crypto}, they represent a good alternative - especially in the last three years - through which criminals engage in illicit activities and sanctions circumvention. This approach will allow Law Enforcement Officers to have a complete picture of criminal \textit{modus operandi}.

\subsubsection{\ackname} This work has been partially supported by the European Union's Horizon 2020 Research and Innovation Program under the project FALCON (Grant Agreement No. 101121281)

%
%
%
\bibliographystyle{splncs04}
\bibliography{btctestbed}
\end{document}